\documentclass[aps, prd,preprintnumbers,nofootinbib, onecolumn]{revtex4}
\usepackage{bm}
\usepackage{latexsym}
\usepackage{dcolumn}
\usepackage{amsmath,amsfonts,amssymb}
\usepackage{graphicx,epsfig}
\usepackage{amsthm}

\newcommand{\be}{\begin{eqnarray}}
\newcommand{\ee}{\end{eqnarray}}
\newcommand{\bea}{\begin{eqnarray}}
\newcommand{\eea}{\end{eqnarray}}

\def\comment#1{}

\newcommand{\lp}{\ell_{\rm p}}
\newcommand{\mpl}{m_{\rm p}}

\usepackage{color}

\definecolor{darkred}{rgb}{.8,0,0}

\definecolor{darkblue}{rgb}{0,0,.7}

\definecolor{darkgreen}{rgb}{0,.7,0}

\begin{document}

%
%
%%%%%%%%%%%%%%%%%%%%%%%%%%%%%%%%%%%%%%%%%%%%%%%%%%%%%%%%%%%%%%
\title{Can an Axion be the Dark Energy particle ? } 
%%%%%%%%%%%%%%%%%%%%%%%%%%%%%%%%%%%%%%%%%%%%%%%%%%%%%%%%%%%%%%
%

%
%
%
%

\author{Elias~C.~Vagenas}\email[email:~]{elias.vagenas@ku.edu.kw}
\affiliation{Theoretical Physics Group, Department of Physics, Kuwait University, P.O. Box 5969, Safat 13060, Kuwait}
%
%
%
%
%\date{\today}
%
%
%
%
%
%
%
\begin{abstract}
\par\noindent
Following a phenomenological analysis done by the late Martin Perl for the detection of the dark energy, 
we show that an axion of energy  $1.5\times 10^{-3}~eV/c^2$ can be a viable candidate for the dark energy particle. 
In particular, we obtain the characteristic length  and frequency of  the axion as a quantum particle.  
Then, employing a relation that connects the energy density with the frequency of a particle, i.e., $\rho\sim f^{4}$, 
we show that the energy density of axions, with the aforesaid value of mass, as  obtained from 
our theoretical analysis is proportional to the dark energy density computed on observational data, 
i.e., $\rho_{a}/\rho_{DE}\sim \mathcal{O}(1)$. 
\end{abstract}
%
%\pacs{03.65.Ta, 03.65-w}
%
\maketitle
%
%
%
%
%%%%%%%%%%%
\section{Introduction}
%%%%%%%%%%%
%
%
%
\par\noindent
One of the most important and still  unsolved problem in contemporary physics is related to the energy content of the Universe. 
According to the recent results of 2015 Planck mission \cite{Ade:2015xua}, the Universe roughly consists of  $69.11 \%$ 
dark energy,  $26.03 \%$ dark matter, and $4.86 \%$ baryonic (ordinary) matter. Therefore, the energy content of 
the dark sector of our Universe is $95 \%$ of the total energy and actually  we do not know anything about 
its nature as well as which are, if any,  the dark energy and matter particles. 
Consequently, we have not directly detect/observe the dark energy and the dark matter.  
%
%
%
%
%
%
%%%%%%%%%%%%%%%%%%%%%%%%
\section{Still no Detection of the Dark Energy ?}
%%%%%%%%%%%%%%%%%%%%%%%%
%
%
%
%
%
\par\noindent
In \cite{Perl:2008qq}, it was stated that it almost comes as a surprise that the dark energy has not been 
directly detected yet since we have detected energy densities which are much less. In particular, the critical energy 
density reads
\be
\rho_{c}=\frac{3 H^{2}_{0}}{8 \pi G} = 7.76 \times 10^{-10} J/m^{3}
\ee 
\par\noindent
with the Hubble constant $H_0$ to be $ (67.74 \pm 0.46)~km\, s^{-1}\,Mpc^{-1}$ \cite{Ade:2015xua} 
and the  Newton's constant $G$ is $6.674\times 10^{-11}~N\, m^{2}\, kg^{-2}$, thus the dark energy density 
becomes
\be
\rho_{DE}=5.36\times 10^{-10} J/m^{3}
\label{dedensity}
\ee
\par\noindent
and the corresponding mass density of the dark energy will be $0.60 \times 10^{-26}~kg/m^{3} $ which is 
equivalently to almost $4 ~protons/m^{3}$. On the other hand, one can do an experiment in a lab with an electric field of 
magnitude $E=1~V/m$ and measure the energy density of the electric field which will be 
\be
\rho_{E}=\frac{1}{2}\epsilon_{0} E^{2}=4.4 \times 10^{-12} J/m^{3}~.
\ee
Therefore, though the energy density of the electric field, i.e., $\rho{E}$ can be detected and measured, 
the dark  energy density, i.e., $\rho_{DE}$,  which is mass less has not been detected yet. Of course, one has to avoid 
to make an experiment for the detection and measurement of the dark energy near the surface of the Earth, or the Sun, 
or the planets, since the energy density of the gravitational field on the Earth's surface is
\be
\rho_G = \frac{1}{8\pi G}g^{2}= 5.7 \times 10^{+10} J/m^{3}
\label{gravaccel}
\ee
with $g$ to be the gravitational acceleration\footnote{The factor in the RHS of equation (\ref{gravaccel}) is obtained by 
comparing the gravitational field on the surface of the Earth, i.e., $g=G M/R^{2}$ with the electric field of a charged 
conducting sphere on its surface, i.e., $E=(1/4\pi\epsilon_{0}) Q/R^{2}$.}.  It is evident that the energy density of the 
gravitational field is much larger than the dark energy density, thus, as already mentioned above,  any attempt for the detection and measurement 
of the dark energy  has to be performed far from the regions of space in which the gravitational field of massive bodies 
is quite strong. 
%
%
%
%
%
%
%%%%%%%%%%%%%%%%%%%%%%%%%%%%%
\section{Properties of a hypothetical dark energy particle}
%%%%%%%%%%%%%%%%%%%%%%%%%%%%%
%
%
%
%
\par\noindent
As already stated, we do not know the nature of the dark energy and we also do not know which is, if any, the dark energy particle.
Making the assumption that such a particle exists, one can proceed with phenomenological arguments in order to obtain 
some properties of this would-be dark energy particle. The starting point will be the quantum mechanical relation between 
the mass and length of a particle\footnote{As expected, this is an equation satisfied by the Planck mass, i.e., 
$\mpl = \sqrt{\frac{\hbar c}{G}}=2.18\times10^{-8}~kg$ and the Planck length,  i.e., 
 $\lp=\sqrt{\frac{\hbar G}{c^{3}}}=1.62 \times 10^{-35}~m$, respectively.}
\be
m \times L = \frac{\hbar}{c}~.
\label{compton}
\ee
\par\noindent
Of course, we are not aware of the mass of the hypothetical dark energy particle, but we have calculated the 
dark energy density, given by equation (\ref{dedensity}), of the dark energy, i.e., $\rho_{DE}$, 
based on observational data. Thus, it will be easy to compute the length scale of the  dark energy particle, 
i.e., $L_{DE}$, and its corresponding frequency, i.e., $f_{DE}$. Therefore, assuming the dark energy particle
is located in the volume $L_{DE}^{3}$,  equation (\ref{compton}) becomes
\be
\rho_{DE} \times L_{DE}^{4}=\hbar c~,
\label{density+freq}
\ee
the length of the dark energy particle reads
\be
L_{DE}=\left(\frac{\hbar c}{\rho_{DE}}  \right)^{1/4}= 88~\mu m
\ee
\par\noindent
and  the corresponding frequency is equal to
\be
f_{DE}=\frac{c}{L_{DE}}=3.40 \times 10^{12}~Hz~.
\label{defrequency}
\ee
%
%
%
%%%%%%%%%%%%%%%%%%%%%%%%%%%%%
\section{Is axion a Candidate for a dark energy particle ?}
%%%%%%%%%%%%%%%%%%%%%%%%%%%%%
%
%
%
%
\par\noindent
If one assumes that the vacuum provides its  energy for the acceleration of the Universe, this means that the 
vacuum energy density, i.e., $\rho_{vac}$, can run for the dark energy density.  Now, one can theoretically compute the 
former using techniques of  Quantum Field Theory while the latter is computed using astronomical observations 
and so given by equation (\ref{dedensity}), and obtain
\be
\frac{\rho_{vac}}{\rho_{DE}}\sim 10^{120}~.
\ee
\par\noindent
This is the well-known ``old" cosmological problem \cite{Burgess:2013ara}. It is obvious that if one would like to suggest a 
candidate for the dark energy particle, this candidate particle, as a starting point,   has to  at least alleviate this ``old" problem. 
At this point, one would like to investigate which are the options for a dark energy particle. As we will see it seems  that axion 
can run for dark energy particle.   Axion is a hypothetical elementary particle and it was suggested as a solution to the strong 
CP problem in QCD \cite{Peccei:1977hh}.  From a theoretical point of view, it is known that the masses of axions 
in the context of QCD   range from  $50~\mu eV/c^2$ to $1500~ \mu eV/c^{2}$ \cite{Borsanyi:2016ksw}.
 In addition, nowadays axions are   also known  as  cold dark matter candidates with masses which range from $50~\mu eV/c^2$ to $200~\mu eV/c^2$ 
\cite{Ringwald:2016yge}. It is clear that there is a range of masses, namely 
$0.2\times 10^{-3}eV/c^2 < m < 1.5\times 10^{-3}eV/c^2$,  which is not ``used" and  it is worth looking into it. 
Therefore, we choose an axion with mass $1.5\times10^{-3}~eV/c^2$ and we will investigate whether it can run for 
dark energy density. Utilizing equation (\ref{compton}), the length of the axion will be
\be
L_{a} = \frac{\hbar}{c\, m_{a}} = 1.32\times 10^{-4}~m
\ee
\par\noindent
while utilizing equation (\ref{defrequency}), the frequency of the axion reads
\be
f_{a}= 2.27 \times 10^{12}~Hz~.
\ee
\par\noindent
At this point, one employs equation (\ref{density+freq})  to obtain a relation that connects the dark energy 
density with the frequency 
\be
\rho_{DE} = \frac{\hbar}{c^{3}} f^{4}_{DE}~.
\ee
It should be stressed that this  equation will also be satisfied by the  energy density of axions, hence
\be
\frac{\rho_{a}}{\rho_{DE}}\sim \left( \frac{f_{a}}{f_{DE}} \right)^{4}\sim \mathcal{O}(1)~.
\ee
\par\noindent
This is really an acceptable and welcomed  result.
%
%
%
%
%
%
%
%
%%%%%%%%%%%%%%
\section{Discussion}
%%%%%%%%%%%%%%
%
%
%
\par\noindent
In this work we have employed a phenomenological analysis presented in Ref. \cite{Perl:2008qq} in order to 
show that an axion of mass $1.5~meV$ is a viable candidate for dark energy particle. In particular, we derived the length 
that is characteristic for the axion as a quantum particle and then obtained its  frequency. Since we produced a relation 
that connects the energy density with the frequency of a particle, i.e., $\rho\sim f^{4}$, we showed that the energy density of 
axions, with the specific value of mass, as  obtained from our theoretical analysis is proportional to the 
dark energy density computed on observational data. 
At this point a  couple of comments are in order. First, since our analysis here is completely phenomenological, it is 
evident that our  acceptable result does not prove that the axion with mass $1.5\times10^{-3}~eV/c^2$  is the dark energy particle. 
However, this analysis can serve as an easy and quick criterion to classify the viable candidates for dark energy particles. 
Second, till now experiments haven't found any evidence for an axion or an axion-like particle (ALP) 
\cite{Anastassopoulos:2017kag}. The  experiment which is looking in the range of masses under consideration here, 
i.e., $m\ge 1~meV$, is  the PVLAS (Polarization of Vacuum LASer) experiment \cite{DellaValle:2015xxa}.
%
%
%
%
%
%
%
%
%%%%%%%%%%%%%%
\section{Acknowledgments}
%%%%%%%%%%%%%%
%
%
%
\par\noindent
I would like to thank the late Nobel Laureate Professor Martin L. Perl for his constructive comments and encouraging me to write 
up a short paper to display this idea. Additionally, many thanks to Federico Della Valle 
for useful correspondences.
%
%
%
%
%
%%%%%%%%%%%%%%%%%

%%%%%%%%%%%%
%
%
%
\end{document}